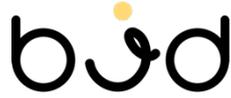

# Inference Acceleration for Large Language Models on CPUs

{jithinvg, dittops, adarsh.ms}@bud.studio

## Abstract

In recent years, large language models have demonstrated remarkable performance across various natural language processing (NLP) tasks. However, deploying these models for real-world applications often requires efficient inference solutions to handle the computational demands. In this paper, we explore the utilization of CPUs for accelerating the inference of large language models. Specifically, we introduce a parallelized approach to enhance throughput by 1) Exploiting the parallel processing capabilities of modern CPU architectures, 2) Batching the inference request. Our evaluation shows the accelerated inference engine gives an **18-22x** improvement in the generated token per sec. The improvement is more with longer sequence and larger models. In addition to this, we can also run multiple workers in the same machine with NUMA node isolation to further improvement in tokens/s. Table 2, we have received **4x** additional improvement with 4 workers. This would also make Gen-AI based products and companies' environment friendly, our estimates shows that CPU usage for Inference could reduce the power consumption of LLMs by **48.9%** (1252 W for A100 with AMD EPYC 7V13 vs 613 W for Intel® Xeon® Gold 6538N) while providing production ready throughput & latency.

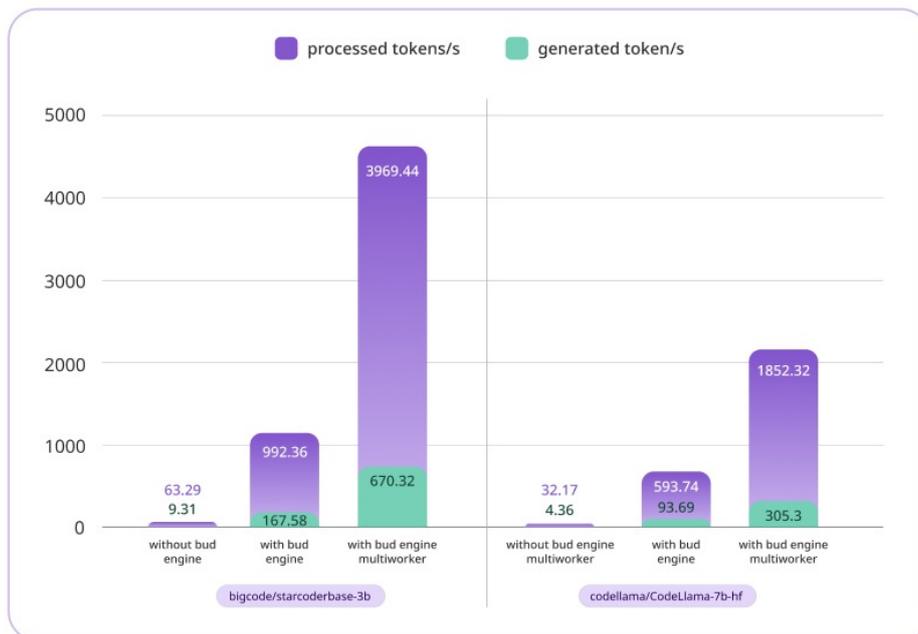

Figure 1: Improvement of token/s with the use of Bud Inference engine with 32 vCPU on 4th Gen Intel® Xeon® Scalable Processors

# 1. Introduction

The widespread integration of large language models (LLMs) across diverse applications, ranging from code generation to writing tasks, has surged in recent times, creating a heightened demand for more efficient inference solutions. Enterprises are increasingly leveraging LLMs to enhance various internal processes. However, the cost associated with their utilization remains a significant concern due to the reliance on GPUs for inference. [1]

The current sequential approach employed by LLMs involves generating one token at a time based on the input prompt and the preceding tokens, persisting until a stop token or the predetermined maximum tokens are reached for each request. This method, while functional, restricts the optimal utilization of available resources [2], [3]. Elevating the tokens generated per second is crucial in mitigating the expenses associated with running LLM-enabled applications. While batching requests can increase throughput, achieving superior batching necessitates optimized memory utilization corresponding to the batch.

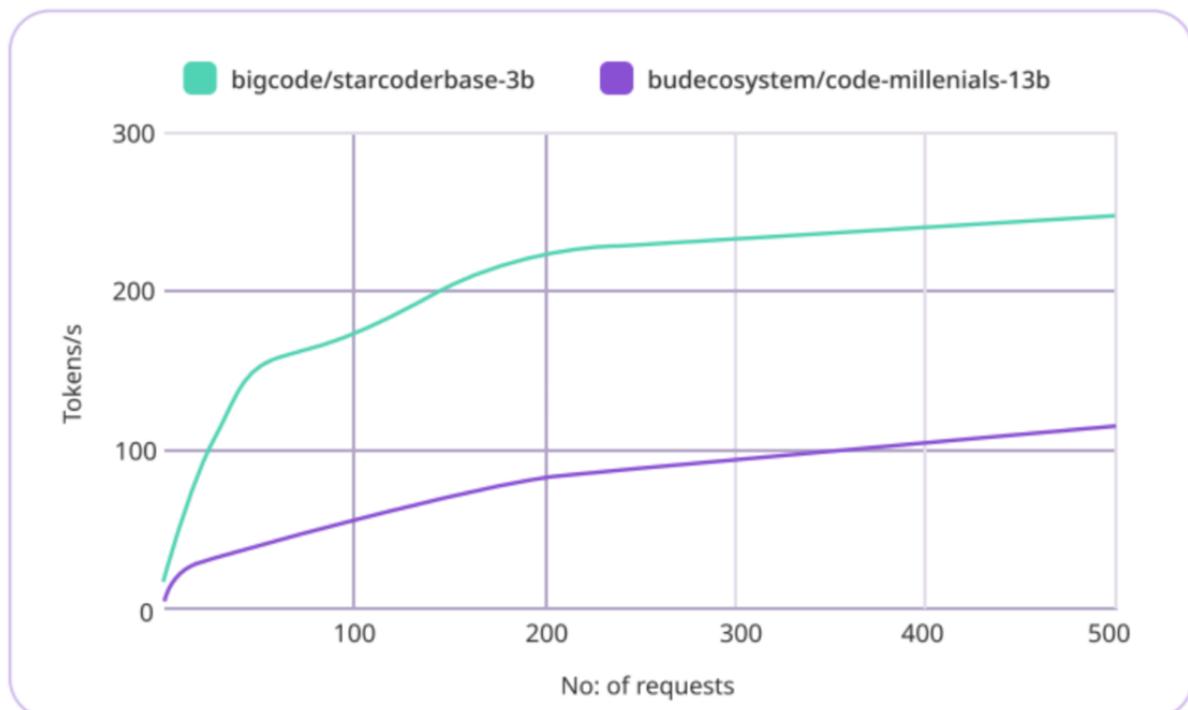

Figure 2: This shows the tokens/s increases with the number of parallel requests increases due to the better utilization of the memory

To address this challenge, we introduce an inference engine designed to enhance memory utilization by partitioning available memory into a series of tiles. This engine efficiently allocates incoming requests and assigns their respective tokens to specific memory tiles based on availability. By overseeing the computation and response delivery, the inference engine optimizes memory allocation per token, significantly enhancing performance for batching multiple requests.

| Model | Number of Request | Without Bud Inference | With Bud Inference |
|---|---|---|---|
| bigcode/starcoderbase-3b | 100 | 9.31 | 167.58 |
| codellama/CodeLlama-7b-hf | 100 | 4.36 | 93.69 |

Table 1: Token generation speed for various models on 4th Gen Intel® Xeon® Scalable Processors with 32 vCPU and 100 requests

In pursuit of cost-effective inference, we explore the feasibility of performing the same computations on Intel® Xeon® Scalable Processors. Leveraging optimizations like AVX and AMX, specifically tailored for Intel® Xeon® Scalable Processors [4], holds the promise of higher throughput in token generation. AVX facilitates parallel execution of operations on broader data vectors, while AMX focuses on optimizing matrix multiplication operations inherent in the core computations of transformer-based language models. The combined effect of these extensions enhances the utilization of the computational power of Intel® Xeon® Scalable Processors, resulting in superior throughput for CPU-based inference tasks.

This paper focuses on harnessing the computational prowess of CPUs to accelerate the inference process. Through the strategic application of parallelization techniques and the utilization of the inference engine for batching, our goal is to enhance throughput, minimize latency, and render large language models more practical for real-time applications.

## 3. Inference Acceleration

Effective memory utilization stands out as a primary challenge in the current CPU-based inference systems. This challenge predominantly arises from the KV (Key-Value) cache, which dynamically adjusts memory utilization in response to the number of tokens [5], [6]. Consequently, the system experiences inefficiencies in memory allocation and grapples with both internal and external CPU memory fragmentation issues. Current systems reserve the required CPU memory for storing the KV cache of a request based on the max token length. But this will lead to internal fragmentation as the actual token generation might be less than the max token length. The reserved memory won't be fully utilized. A request with a smaller token length won't be able to utilize this unused reserved memory. Also, the max length for different request would be different which lead to external memory fragmentation. This limitation is there in GPU inference systems as well, as mentioned in paged attention [7].

Memory sharing could be useful for batching simultaneously request effectively. But memory sharing is not possible in the current systems. The KV cache of each request has its own contiguous space which isolate each request in the CPU memory.

As suggested in paged attention, one of the solutions of resolve these inefficiencies is to create memory tiles which then can be utilized efficiently based on the request. So, to streamline the inference process, the memory manager within the inference engine strategically divides the available CPU memory into a set of n tiles. The memory manager indexes these tiles with the physical CPU memory. The request's KV cache will be divided into smaller chunks and allocated to specific memory tiles based on the availability in the index. This avoids the necessity of contiguous space requirement. The engine also takes multiple requests and fill memory tiles based on the availability. This ensures that the reserved memory is fully utilised when there are enough request coming to the engine. When a request is completed, the KV cache memory is released back to the index so that it can be utilized for the next request.

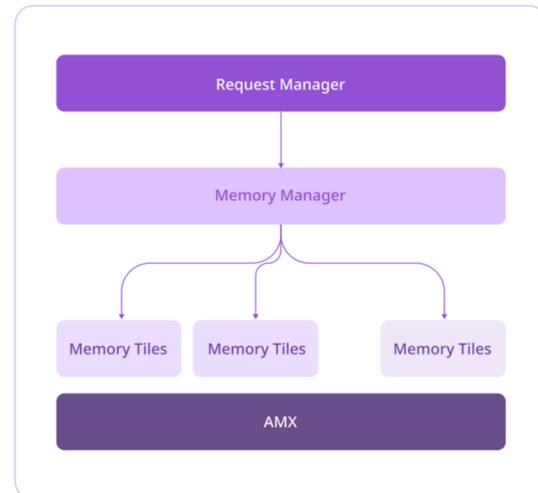

Figure 3: Memory allocation in memory engine

Intel® AMX enables AI workloads to run on the CPU instead of offloading them to a discrete accelerator, providing a significant performance boost [4]. Its architecture supports BF16 and int8 data types and includes two main components:

- Tiles: These consist of eight two-dimensional registers, each 1 kilobyte in size, that store large chunks of data.
- Tile Matrix Multiplication (TMUL): TMUL is an accelerator engine attached to the tiles that performs matrix-multiply computations for AI.

Together, these components enable Intel® AMX to store more data in each core and compute larger matrices in a single operation. Additionally, Intel® AMX is architected to be fully extensible and scalable. The inference engine utilizes tiles and TMUL for faster matrix calculations in the model architecture. This gives additional boost in the Intel® Xeon® Scalable Processors for the token generation.

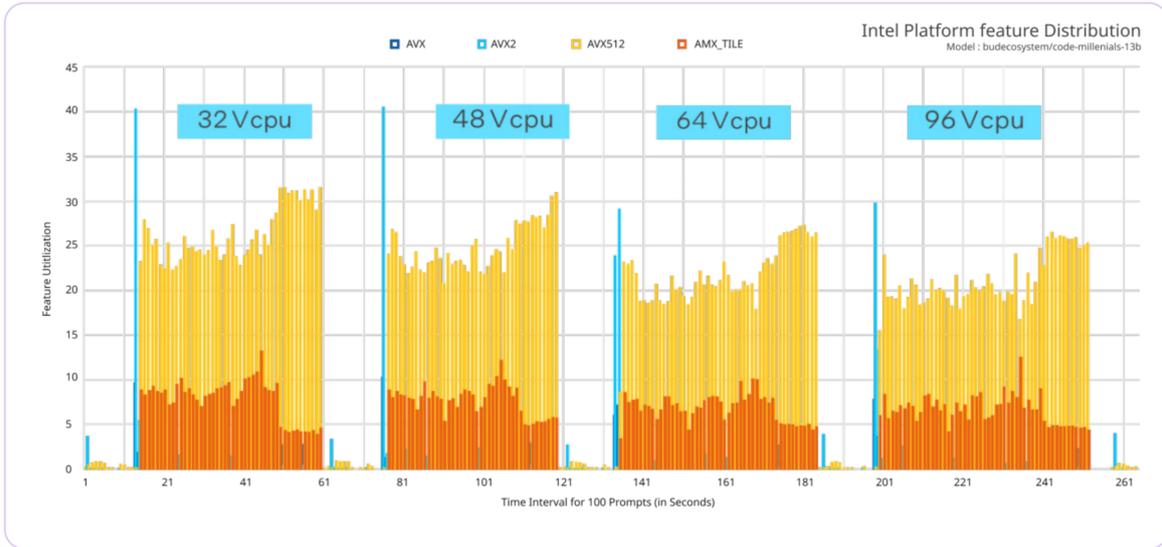

Figure 4: feature utilization of AVX/AMX in the 4[th] Gen Intel® Xeon® Scalable Processor

The throughput per machine can further improved by optimizing the CPU utilization with multiple workers. Isolating a process to run on cores specific to a NUMA node will give higher performance by reducing memory latency [8]. The engine constrains each worker to run on a specific NUMA node. For a machine with 4 NUMA node, 4 different workers with isolated cores for each process will be initiated. This will give higher throughput compared to process running across multiple NUMA nodes. We have also found that, setting the number of threads used for each worker with slightly lesser than the number of cores in a NUMA node will give optimal performance. Table 2, By defining the cores and threads, we were able to run 4 workers in a machine and get an overall throughput of 1852.32 processed tokens/sec and 305.30 generated token/s for LLaMA 7B.

| Model | Node | Number of requests | Execution time(mm:ss) | Processed Token Throughput (Tokens/Sec) | Generated Token Throughput (Tokens/Sec) |
|---|---|---|---|---|---|
| meta-llama/Llama-2-7b-hf | NUMA 0 | 1000 | 598:19 | 466.56 | 76.90 |
| meta-llama/Llama-2-7b-hf | NUMA 1 | 1000 | 599:57 | 465.49 | 76.72 |
| meta-llama/Llama-2-7b-hf | NUMA 2 | 1000 | 605:91 | 460.61 | 75.92 |
| meta-llama/Llama-2-7b-hf | NUMA 3 | 1000 | 607:18 | 459.66 | 75.76 |
| | | | **Aggregate per server** | **1852.32** | **305.30** |

Table 2: Token generation on 2 x Intel® Xeon® PLATINUM 8592+ with 4 workers

## 4. Experimental Setup

We used starcoder 3B & 7B [9], codellama 7B & 13B [10] and code-millenials 13B, 34B models for the evaluation of the engine. We have used 4th Gen Intel® Xeon® Scalable Processors with 32 & 48 vCPU.

The experiment is run on a set of instructions data to simulate parallel request and evaluate the performance of the engine.

## 5. Results and Analysis:

In this section, we present the results of our experiments, showcasing the performance gains achieved through CPU utilization. We analyse the impact of different parallelization techniques on throughput, latency, and resource utilization. Additionally, we discuss any trade-offs or limitations observed during experimentation.

Table 3 shows the performance of the engine on 4th Gen Intel® Xeon® Scalable Processors. The engine was tested on 32vCPU and 100 parallel requests. The inference engine and AMX add 18x improvement in token generation per sec on *bigcode/starcoderbase-3b* as shown in Figure 1.

| vCPU | Model | Number of requests | Processed Token Throughput (Tokens/Sec) | Generated Token Throughput (Tokens/Sec) | Execution time(mm:ss) |
|---|---|---|---|---|---|
| 32 | bigcode/starcoderbase-3b | 100 | 992.36 | 167.58 | 0:27 |
| 32 | bigcode/starcoderbase-7b | 100 | 506.39 | 85.52 | 0:53 |
| 32 | bigcode/starcoderbase-15b | 100 | 273.92 | 46.26 | 1:39 |
| 32 | budecosystem/code-millenials-13b | 100 | 321.81 | 50.57 | 1:30 |
| 32 | budecosystem/code-millenials-34b | 100 | 149.3 | 23.46 | 3:16 |
| 32 | codellama/CodeLlama-7b-hf | 100 | 593.74 | 93.69 | 0:49 |
| 32 | codellama/CodeLlama-13b-hf | 100 | 341.55 | 53.89 | 1:25 |

Table 3: Token generation speed for various models on 4th Gen Intel® Xeon® Scalable Processor with 32 vCPU and 100 requests

The experiments also show increase in the token generation per sec as we increase the number of vCPU. So, the engine can vertically scale based on the available vCPUs. But as shown in Figure 2, the better utilization happens based on the number of requests as well. As the number of requests increases, the token generation per sec also increases. For larger vCPUs we would need higher number of requests to fully utilize the resource.

| vCPU | Model | Number of Prompts | Processed Token Throughput (Tokens/Sec) | Generated Token Throughput (Tokens/Sec) | Execution time(mm:ss) |
|---|---|---|---|---|---|
| 48 | bigcode/starcoderbase-3b | 100 | 1075.19 | 181.57 | 0:25 |
| 48 | bigcode/starcoderbase-7b | 100 | 445.55 | 75.24 | 1:01 |
| 48 | bigcode/starcoderbase-15b | 100 | 286.69 | 48.42 | 1:35 |
| 48 | budecosystem/code-millenials-13b | 100 | 333.11 | 52.35 | 1:27 |
| 48 | budecosystem/code-millenials-34b | 100 | 157.63 | 24.77 | 3:05 |
| 48 | codellama/CodeLlama-7b-hf | 100 | 536.59 | 84.67 | 0:54 |
| 48 | codellama/CodeLlama-13b-hf | 100 | 372.77 | 58.82 | 1:18 |

Table 4: Token generation speed for various models on 4th Gen Intel® Xeon® Scalable Processors with 48 vCPU and 100 requests

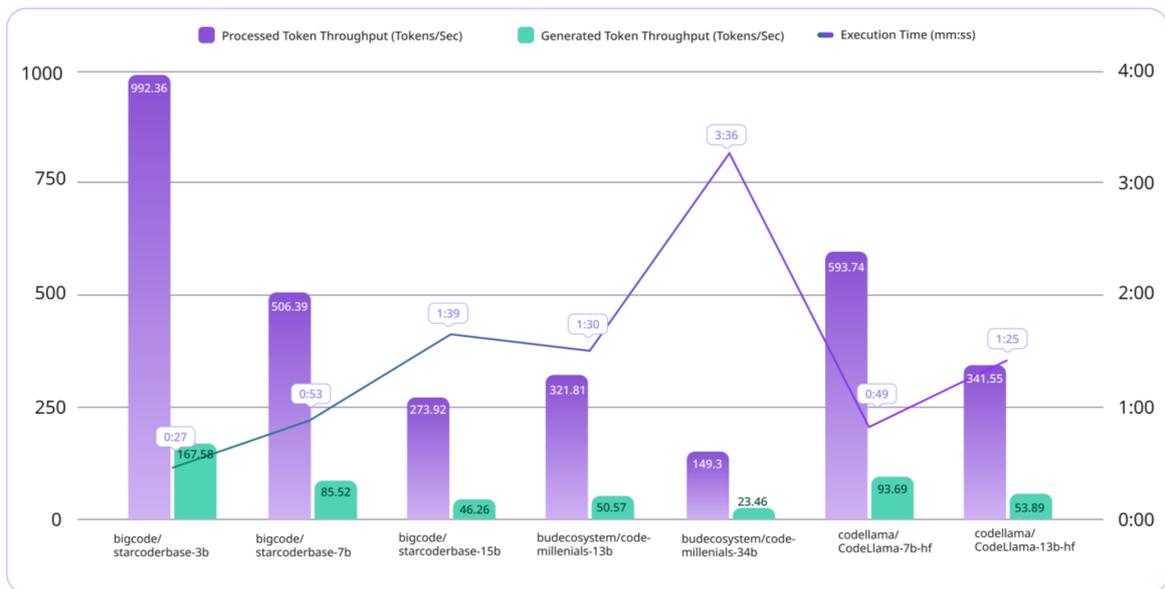

Figure 5: Comparison of different models on 4th Gen Intel® Xeon® Scalable Processors with 32 VCPU. The figure shows how the processed token/s and generated token/s vary based on the model parameter size

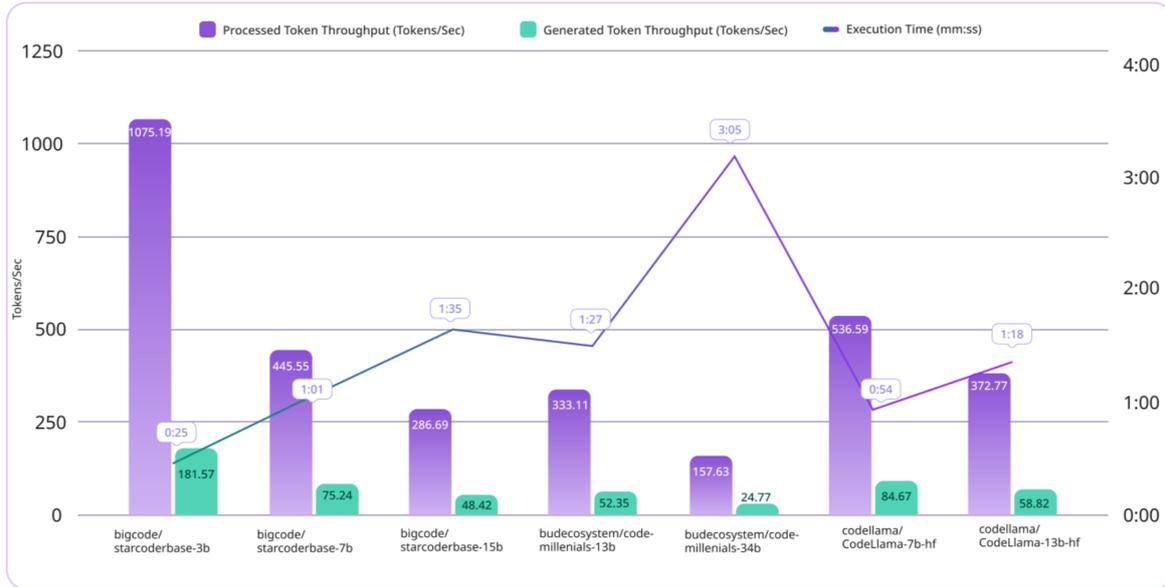

Figure 6: Comparison of different models on 4th Gen Intel® Xeon® Scalable Processors with 48 VCPU. The figure shows how the processed token/s and generated token/s vary based on the model parameter size

| Machine | CPU | GPU | Total Power | Token/s | Power for 1K tokens |
|---|---|---|---|---|---|
| Azure Standard_NC24ads_A100_v4 | AMD EPYC 7V13 | A100 80GB | 640 W | 511 | 1252 |
| Intel® Xeon® Gold 6538N Processor | 2 * Intel® Xeon® Gold 6538N Processor | - | 410W | 668 | 613 |

Table 5: Power consumption comparison between CPU and GPU for bigcode/starcoderbase-3b

## 6. Conclusion

In conclusion, we demonstrate the effectiveness of parallelized inference on CPUs for large language models. Our approach provides a scalable solution to enhance throughput, making these models more practical for deployment in real-world applications. We discuss potential avenues for future research and optimizations to further improve the efficiency of large language model inference on CPUs.